\documentclass[conference]{IEEEtran}
\usepackage{ifthen,color}
\definecolor{darkgreen}{rgb} {0.0,0.5,0.0}
\definecolor{darkblue}{rgb} {0.0,0.0,0.5}
\definecolor{bluegreen}{rgb}{0.0,0.5,0.5}

\newtheorem{definition}{Definition}
\usepackage{placeins}
\usepackage{microtype}
\usepackage[export]{adjustbox}
\usepackage{enumitem}
\usepackage{pdfpages}
\usepackage{verbatim}
\DeclareMathAlphabet{\mathcal}{OMS}{cmsy}{m}{n}
\SetMathAlphabet{\mathcal}{bold}{OMS}{cmsy}{b}{n}
\usepackage{multirow}
\usepackage{booktabs}

\usepackage{cite}

\usepackage{graphicx}
\usepackage{amsmath}
\usepackage{algorithmicx,algorithm,algpseudocode}
\ifCLASSOPTIONcompsoc
  \usepackage[caption=false,font=normalsize,labelfont=sf,textfont=sf]{subfig}
\else
  \usepackage[caption=false,font=footnotesize]{subfig}
\fi
\usepackage{url}
\usepackage{multirow}

\begin{document}
%
\title{Fast BFS-Based Triangle Counting on GPUs}

\author{\IEEEauthorblockN{Leyuan Wang}
\IEEEauthorblockA{Department of Computer Science\\
University of California, Davis\\
Davis, California 95616\\
Email: leywang@ucdavis.edu}
\and
\IEEEauthorblockN{John D. Owens}
\IEEEauthorblockA{Department of Electrical \& Computer Engineering\\
University of California, Davis\\
Davis, California 95616\\
Email: jowens@ece.ucdavis.edu}}


%


\maketitle

\begin{abstract}
In this paper, we propose a novel method to compute triangle counting on GPUs. Unlike previous formulations of graph matching, our approach is BFS-based by traversing the graph in an all-source-BFS manner and thus can be mapped onto GPUs in a massively parallel fashion. Our implementation uses the Gunrock programming model and we evaluate our implementation in runtime and memory consumption compared with previous state-of-the-art work. We sustain a peak traversed-edges-per-second (TEPS) rate of nearly 10 GTEPS\@. Our algorithm is the most scalable and parallel among all existing GPU implementations and also outperforms all existing CPU distributed implementations. This work specifically focuses on leveraging our implementation on the triangle counting problem for the Subgraph Isomorphism Graph Challenge 2019, demonstrating a geometric mean speedup over the 2018 champion of 3.84$\times$.
\end{abstract}

%
\IEEEpeerreviewmaketitle

\section{Introduction}
\label{sec:intro}
Previous work in Graph Challenge has explored triangle counting methods, including set-intersection and matrix-formulation approaches based on sparse matrix-matrix multiplications. But none of them has dived deep into the graph matching approach. In our previous work~\cite{Wang:2016:GAH}, we did a comparative study among set intersection, matrix multiplication and graph matching approaches, and found great potential in a graph matching approach. In this paper, we address the triangle counting problem with a BFS-based subgraph matching to a triangle pattern.

Graph matching maps one graph onto another in such a way that both the topological structure and the node and edge labels are matched between the two graphs. The problem can be formalized as the search for all subgraph isomorphisms between two graphs. Most previous subgraph isomorphism algorithms fall into the following three classes: depth-first tree search, constraint propagation, and graph indexing. However, none of them is efficient on GPUs (Section~\ref{sec:related}). Prior work on GPUs only targets specific applications and is generally memory-bounded.

Many graph libraries make attempts to solve some common operations on parallel machines, including the Parallel Boost Graph Library (PBGL), Pregel~\cite{Malewicz:2010:PSL}, GraphLab~\cite{Low:2010:GAN}, PowerGraph~\cite{Gonzalez:2012:PDG}, Ligra~\cite{Shun:2013:LAL}, Gunrock~\cite{Wang:2017:GGG}, and nvGraph.\footnote{nvGRAPH is available at \url{https://developer.nvidia.com/nvgraph/}.} Modern graphics processors (GPUs) have been leveraged in those operations and have found success in several parallel applications. But few of those libraries have efficiently solved the subgraph isomorphism problem, either because the framework is not fit for the problem or because no method fully exploits the large amount of parallelism available on a GPU\@. Existing solutions are mostly based on either backtracking or filtering-and-joining mechanisms. Backtracking-based methods do not fit well on GPU architectures because of their recursive nature. Existing filtering-and-joining strategies that focus on optimizing matching order either generate a large amount of intermediate results or handle intermediate results in an inefficient way. Compared with other triangle counting methods such as set intersection, which avoids the problem of generating intermediate results, filtering-and-joining approach has disadvantages.

Our work effectively uses the compute power of an entire GPU with an approach that leverages an an existing GPU high-performance graph processing framework as well as existing high-performing GPU computing primitives. Our approach scales especially well because we optimize our algorithm to make it generate as few intermediate results as possible.

The paper is organized as follows: section~\ref{sec:related} formalizes the problem, compares with previous methods in solving the problem, and introduces the graph framework we leverage in the implementation. Section~\ref{sec:approach} describes our BFS-based graph matching method in detail, and illustrates how we uses it to solve triangle counting. Section~\ref{sec:exp} summarizes our experimental results. And section~\ref{sec:conc} concludes the paper with takeaways as well as future work.

\section{Preliminaries and Related Work}
\label{sec:related}
\subsection{Problem Definition}
Triangle counting is the problem of finding all occurrences of a triangle in a  graph. The occurrences should match the original triangle graph both structurally and semantically. In other words, both the graph topology and attribute information for nodes and edges should be considered when determining the similarity between two graphs. In the Graph Challenge problem, both graphs are undirected with no node labels or edge weights. We use the following mathematical definitions:

\begin{definition}
A graph is a 3-tuple $G=(V,E)$, where $V$ is a set of vertices, $E \subseteq V \times V$ are the edges connecting those vertices.
\end{definition}

We represent the vertex and edge set of the graph $G$ respectively as $V(G)$ and $E(G)$. If two vertices in $G$, say $u,v \in V$, are connected by an edge $e \in E$, which is denoted by $e=(u, v)$, then $u,v$ are \textit{adjacent} or \textit{neighbors}. A graph is \textit{undirected} when it only contains edges that have no direction, meaning $(u,v)$ and $(v,u)$ essentially represent the same edge. Though this paper only describes the problem on undirected graphs, it can be extended to directed ones easily. With these preliminaries, we define subgraph isomorphism as follows:
\begin{definition}
A graph $G = (V, E)$ is subgraph-isomorphic to another graph $G' = (V', E')$, denoted as $G \subseteq G'$, if there is an injection function $f:V\rightarrow V'$, such that
\begin{center}
    $\forall (u,v) \in E: (f(u), f(v)) \in E'$.
\end{center}
\end{definition}

Given a triangle as a query graph $Q$ and a data graph $G$, the exact triangle counting problem enumerates all triangles that are isomorphic to $Q$ in $G$. So the inputs we take are one query graph, $Q$, which is a triangle, and one data graph, $G$, in MatrixMarket format. Our outputs are the number of matches as well as the matched subgraph node ID lists from the data graph $G$. Compared with other methods such as set intersection and matrix multiplication, one advantage of using subgraph matching to solve triangle counting is that we can get the triangle listings for free. Another advantage is that our implementation could potentially be extended to embeddings other than triangles with/without node and/or edge label information.

\subsection{The Gunrock graph processing framework}
\begin{figure}[ht]
\begin{center}
\includegraphics[width=\columnwidth]{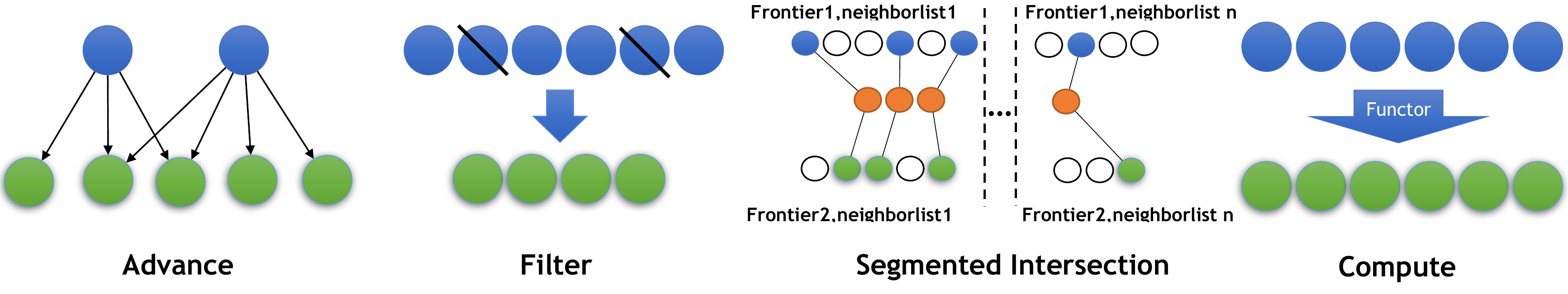}
\end{center}
\caption{Gunrock framework graph traversal operators.\label{fig:gunrock}}
\end{figure}

We note several graph frameworks in Section~\ref{sec:intro}; in this work, we choose the Gunrock~\cite{Wang:2017:GGG} GPU-based graph analytics framework. Gunrock uses a high-level, bulk-synchronous, data-centric abstraction. Gunrock programs are expressed as manipulations of frontiers of vertices or edges that are actively participating in the computation. Its traversal-based operators (shown in Fig.~\ref{fig:gunrock}) currently include:
\begin{description}
  \item[Advance] which generates a new frontier via visiting the neighboring vertices/edges in the current frontier (work distribution/load balancing).
  \item[Filter] which removes elements from a frontier via validation tests.
  \item[Segmented intersection] which computes the intersection of two neighbor lists for each pair of elements from two input frontiers.
  \item[Compute] user-defined vertex/edge-centric computations that run in parallel; they can be combined with advance or filter.
\end{description}
The Gunrock framework is very efficient for BFS-based algorithms. Since our algorithm is also BFS-based, we use the Gunrock framework to fully utilize the massive parallelism of GPUs by using the above four operators. We also leverage Gunrock's capability of supporting frontiers of either nodes or edges. We describe the pros and cons of using Gunrock in Section~\ref{sec:approach}.

\section{Approach}
\label{sec:approach}
Most existing approaches follow either a \emph{filtering}-and-\emph{verification} or \emph{filtering}-and-\emph{joining} strategy. The filtering step prunes out candidates that cannot contribute to the final solutions; we later show why this step determines the efficiency of the algorithm. The verification step is generally based on Ullman's backtracking subroutine~\cite{Ullmann:1976:ASI} which searches in a depth-first manner for matchings between the query graph and the updated data graph obtained from the filtering step. The joining step combines the filtered candidate edges or partial results to find all matches. Previous graph-index-based methods mostly follow  \emph{filtering}-and-\emph{joining}, which was also our approach when we initially designed our algorithm. But we found that the intermediate results generated before the \emph{joining} step is a challenge to store on a single GPU\@. Instead, the optimized method that we propose in this paper follows a \emph{filtering}-and-\emph{verification} approach, but unlike previous work, it is not based on depth-first search.

The aim of the filtering step is to reduce the search space on which later verification or joining steps operate. A good filtering technique can save significant effort by pruning out non-valid nodes/edges before verification or joining, which is usually the bottleneck of the whole algorithm. Filtering mechanisms can be classified into two categories depending on their exploring scopes: local or global. A local refinement mechanism prunes the set of mappings that are candidates for each single vertex. A global pruning reduces the global search space. In our approach, we use an effective filtering method by neighborhood footprint encoding which encodes each node's neighborhood information. The encoding information is updated after each local pruning and thus generates a more effective global pruning of the search space.

We traverse the data graph in an all-source-BFS manner. In other words, we start BFS traversals starting from every vertex in the data graph in parallel. The verification step happens in every iteration when we decide whether to add the newly visited node to the partial results; the verification test itself is based on the previously stored constraints. In order to avoid an explosion in memory requirements, we do a compaction after each iteration to prune out invalid partial results. In this way, we are able to keep the memory usage linear to the number of the number of matched subgraphs in the data graph.

In this section, we illustrate our approach step by step.

\subsection{The proposed algorithm}
\label{subsec:algorithm}
The algorithm we propose achieves the following goals:
\begin{enumerate}
\item Scalable on GPU cores.
\item Memory consumption proportional to the number of matched triangles.
\end{enumerate}

Unlike most previous methods that are based on either tree search or graph indexing, our algorithm is BFS-based, which is a better fit for the GPU\@. In terms of memory usage, though the algorithm is not recursive, we are still able to limit the space needed to space proportional to the number of edges in the graph by using more efficient pruning techniques and selecting query order in a novel way. Algorithm~\ref{alg:tcsm} is the pseudocode that shows both the algorithm itself and its implementation using the Gunrock framework. Our inputs include a triangle as a query graph $Q$ and a large data graph $G$ for searching. Both graphs are undirected. Our output includes the exact number of matched triangles as well as the node sequence listings of them.

\begin{algorithm}[!ht]
\label{alg:tcsm}
\renewcommand{\algorithmicrequire}{\textbf{Input:}}
\renewcommand{\algorithmicensure}{\textbf{Output:}}
        \begin{small}
                \begin{algorithmic}[1]
                  \Require{Triangle Graph $Q$, Data Graph $G$.}
           \Ensure{Count of triangles $n$ and listings of all matched triangles.}
                      \Procedure{PreCompute\_on\_CPUs}{}
                                \State
                                                             \Call{Store\_none\_tree\_connection}{$E$}
                                 \State
                                 \Call{Generate\_UMO}{NEC}
                                 \EndProcedure
           \Procedure{Filtering\_candidate\_set}{$Q,G$}
                        \State
                        \Call{Advance+Compute}{$G$}\Comment{Compute NE for each node}
                        \State
                        \Call{Filter+Compute}{$G,Q,c\_set$}\Comment{Filter nodes based on (NE); update NE}
                        \EndProcedure
                        \While{$(\lvert M[i]\rvert < \lvert Q\rvert)$}
                        \Procedure{Verifying\_Constraints}{$G,Q,c\_set, M$}
                        \State
                        \Call{Advance}{$c\_set$}\Comment{BFS traversal from source nodes in $c\_set$ to dest nodes that are verified on stored constraints.}
                        \State
                        \Call{Compute}{$c\_set$}\Comment{Compact satisfied dest nodes to $c\_set$.}
                        \State
                        \Call{Write\_to\_Partial}{$M$}\Comment{Add updated $c\_set$ to partial results $M$.}
                        \State
                        \Call{Mask}{$M$}\Comment{Set incomplete partial results to invalid values.}
                        \EndProcedure
                        \EndWhile
                        \State
                        \Return{Triangle count:$\frac{\lvert M\rvert}{\lvert Q\rvert}$, $M$}
                \end{algorithmic}
        \end{small}
\end{algorithm}

For the pre-computing part, need to store the query node connection information. We make sure that the node sequence we generate meets the BFS traversal of a spanning tree derived from $Q$. So we need to store any non-tree edge connection (line 2 of Alg.~\ref{alg:tcsm}). We maintain a set of constraints on query node ID values to avoid generating duplicated subgraphs. We call this a \emph{unique mapping order} (UMO) (line 3 of Alg.~\ref{alg:tcsm}). The concept is derived from the neighborhood equivalence idea from Turbo$_{\text{ISO}}$~\cite{Han:2013:TIT}.
\begin{definition}
Any pair of nodes $u_i, u_j \in V(Q)$ are neighborhood-equivalent (denoted by $\simeq$), if for every embedding $m$ that contains node mapping pairs $(u_i, v_i)$ and $(u_j, v_j)$ where $v_i, v_j \in V(G)$, there exits an embedding $m'$ such that $m'=m - \{(u_i, v_i), (u_j, v_j)\} \cup \{(u_i, v_j), (u_j, v_i)\}$.
\end{definition}

The above definition illustrates that neighborhood equivalent query nodes share the same matched vertices in data graph. The equivalence class of a query vertex $u$ is a set of query vertices that are neighborhood-equivalent to ($\simeq$) $u$. This class is called the neighborhood equivalence class (NEC). The Turbo$_{\text{ISO}}$~\cite{Han:2013:TTU}  paper proves a lemma that, in an NEC, for each node $u\in V(Q)$, either an NEC member $u_n$ has the same label and same set of adjacent vertices; or every member of the NEC has the same label and are adjacent to each other. By leveraging the NEC idea, we first find NECs in the query graph, and inside each NEC, we define an ordering based on node IDs and store the orderings as constraints. For example, a triangle query graph itself is an NEC. The ordering we define is $\{u_1, u_2, u_3\}$ ($u_1,u_2,u_3$ are node IDs of the query graph) where $u_1<u_2<u_3$. So when doing triangle matching in the data graph, we only traverse edges with a destination node ID value larger than the source node ID value.

The main algorithm can be separated into two big steps: \emph{filtering} and \emph{verification}. The filtering step starts with a computation of neighborhood encoding (NE) (line 6 of Alg.~\ref{alg:tcsm}), which is computed based on the degrees of nodes in the data graph. Here a node $u$'s NE is defined as the degree of $u$. NE is based on an idea of using an integer to represent neighborhood information that characterizes each vertex in the data graph. We compute each triangle node's NE during pre-processing on the CPU and filter the candidate nodes in $G$ based on NE (line 7 of Alg.~\ref{alg:tcsm}). Note that NE information is updated once we filter out non-valid candidate nodes. The verification step is based on multi-source breadth-first search, where the sources are the candidate set of nodes from the filtering step.  After each traversal, we verify if the destination nodes can be added to the partial results based on the pre-computed constraints (line 11 of Alg.~\ref{alg:tcsm}). Note that during verification, we avoid excessive memory usage by doing compaction before writing to partial results (line 13 of Alg.~\ref{alg:tcsm}),  and before the next traversal, we mask out the unfruitful partial results that contain nodes/edges less than the current stage number (line 14 of Alg.~\ref{alg:tcsm}). The number of parallel traversals depends on the level of the query graph spanning tree. So in the pseudocode of Alg.~\ref{alg:tcsm}, we use a while loop to ensure partial results are fully complete before returning the final results.

\subsection{Implementation}
\begin{figure}[ht]
\begin{center}
\includegraphics[width=0.85\columnwidth]{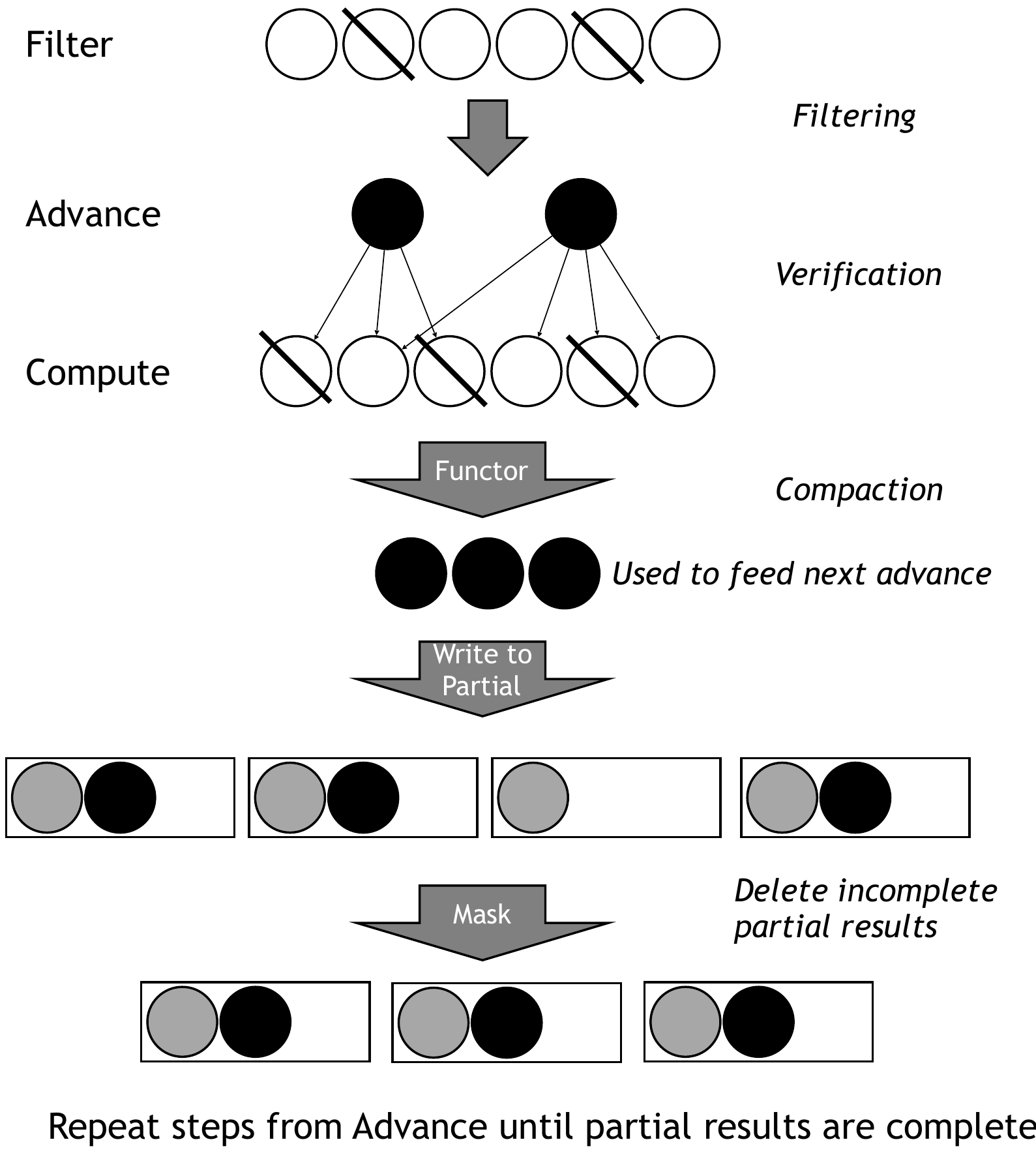}
\caption{Implementation flow chart. \label{fig:sm}}
\end{center}
\end{figure}

To better understand Alg.~\ref{alg:tcsm}, we draw an implementation flowchart, which can be found in Figure~\ref{fig:sm}.
We use compressed sparse row (CSR) format as our data structure to store graphs in a space-efficient fashion.
Filter, advance and compute are the three operators we use from Gunrock library. The filter operator takes in all nodes from the data graph $G$ and returns nodes with satisfied NE requirements. Next, we use the advance operator to traverse all the neighbor nodes of the candidate nodes. So we need to verify if the newly traversed edges in the data graph satisfy connection constraints with corresponding query edges. The connection constraints include both the connection with existing partial results as well as requirements brought by corresponding query edges. By using the advance operator, this step is done in a massively parallel manner where the large amount of newly added edges are mapped to consecutive GPU threads and the verification computation is also done in a SIMT manner. Then we get output neighbor nodes from the threads that pass the constraint-verification tests of the advance operator. In the next step, we use a compute operator to compact the candidate nodes from scattered threads to consecutive positions in order to serve as the inputs for the next iteration of advance traversal. Note that in GPU computing, consecutive (coalesced) reads perform much better than scattered reads. So not only do we benefit from better memory complexity but also gain a performance boost from compaction.

\subsection{Optimizations}

Our first optimization is $k$-look-ahead, borrowed from the VF3 algorithm, in the verification step to filter out more redundant partial results in each iteration according to feasibility rules besides using connection constraints.
$k$-look-ahead originates from the idea that it is possible to prove a non-consistent state will not generate any successive consistent states. However, it may happen that, even if the state is consistent according to our constraints verification, after a certain number of steps it cannot generate any consistent descendants and, thus, cannot contribute to the final results. The detection of such situations would help to further reduce the number of explored states by avoiding the generation of consistent, but unfruitful, states. In order to achieve this aim, the algorithm verifies if the addition of the new candidate nodes to the partial results generates a consistent state; in addition, it is able to detect $k$ steps in advance if the state will not have any consistent descendants, a $k$-look-ahead.
Note that this optimization is a necessary but not sufficient condition. If it is false, it guarantees that the current candidate node will not pass the next iteration of verification. In our implementation, we use 1- and 2-look-ahead only, which in practice prunes out a significant fraction of unfruitful partial results.

Another optimization is our method of avoiding duplicated final results that cannot be discovered when they are incomplete. For instance, if we visit a certain part of the data graph with a different node sequence, we can get different partial results at the beginning but the same solution in the end, which increases the number of intermediate results and requires further effort to filter out duplicated final results as well.  One possible solution is to use a hash table to keep track of all node sequences. Though it stops the generation of duplicate results, it cannot filter out unfruitful intermediate results. Moreover, such a large hash table storing all partial solutions would be very hard to fit on a single GPU\@. Instead we add constraints to the query node visiting order to ensure that no duplicate final results would be generated from the beginning. We successfully transfer the idea of a \emph{node equivalence class} (NEC), which was previously used to reduce a depth-first search space to solve this problem (mentioned in Section~\ref{subsec:algorithm}). For example, consider a triangle: all three nodes of a triangle are equivalent. So we need to define a visiting order of those equivalent nodes to avoid visiting the same combination multiple times. So in this example, we relate the visiting order to node id value and only visit in the order of increasing node id values. In this way, the same combination of nodes will only be visited once in the data graph.

As a summary, we use the following optimizations:
\begin{enumerate}
\item  Compaction used after each advance operation to move the scatterly distributed intermediate results to adjacent spaces in memory, which improves data access efficiency as well as memory usage.
\item $k$-look-ahead before adding a new node to partial results to prune out unfruitful intermediate results.
\item Encoding neighborhood information for better filtering and constraint verification.
\item Use node equivalence idea to avoid generating duplicated solutions and redundant intermediate results.
\item Mask out incomplete partial results at the end of each iteration to save memory usage.
\end{enumerate}

\section{Experiments and Results}
\label{sec:exp}
\subsection{Experimental Setup}
\paragraph{System}
We tested our triangle counting implementation on an NVIDIA Titan~V GPU. The Titan~V is a Volta-based GPU with 80 streaming multiprocessors (SMs) and 12 GB HBM2 memory. The total memory bandwidth of the Titan~V is 652.8 GB/s. In our experiments, we compare with last year's champion~\cite{Hu:2018:HTC}. In their paper, they run experiments on an NVIDIA Tesla P100 GPU from the San Diego Super Computer Center (SDSC)\@.  This P100 GPU is a Pascal-based GPU with 56 SMs and 16 GB CoWoS HBM2 memory at 732 GB/s memory bandwidth. Our implementation is compiled with CUDA 10.0.
\paragraph{Dataset}
The experiments are done using both real-world and synthetic datasets from the HPEC graph challenge. The graphs are unlabeled graphs. Specifically, we compare results generated from small graphs recognized by Hu et al.~\cite{Hu:2018:HTC} for the reason that our implementation currently only supports one GPU and the GPU memory is limited. But our implementation could be extended to efficient multi-GPU implementation easily under the Gunrock framework. We also assume the input graph is undirected in our implementation.
\subsection{Results}
In Table~\ref{tab:perf}, we show our performance in time and transacted edges per second (TEPS)\@. We compare with last year's champion~\cite{Hu:2018:HTC}. From the table we can see that we are consistently faster than Hu et al.~\cite{Hu:2018:HTC} in all real-world datasets, but slower for one synthetic dataset. Note that the given synthetic datasets have a larger number of triangles, which means more intermediate results will be generated during the computation and thus slow down our implementation.
\begin{table*}[tbph]
\centering
\footnotesize
\begin{tabular}{@{}lcccccc@{}}
\toprule
Graph     & $\lvert V \rvert$ & $\lvert E \rvert$   & Triangles    & Runtime (ms) & Rate (TEPS) & Speedup \\
\midrule
amazon0302 & 262,112 & 899,792  & 717,719 &                     0.445414        &       2.02E+09        &       9.53    \\
amazon0312 & 400,728 & 2,349,869 & 3,686,467 &  2.707648        &       8.68E+08        &       4.23    \\
amazon0505 & 410,237 & 2,439,437 & 3,951,063 &  2.840805        &       8.59E+08        &       4.17    \\
amazon0601 & 403,395 & 2,443,408 & 3,986,507 &  2.836609        &       8.61E+08        &       3.73    \\
as20000102 & 6,475  & 12,572 & 6,584 &          0.298905        &       4.21E+07        &       1.01    \\
as-caida20071105 & 26,476 & 53,381 & 36,365 &   0.599098        &       8.91E+07        &       8.17    \\
ca-AstroPh & 18,773 & 198,050 & 1,351,441 &     0.427914        &       4.63E+08        &       4.94    \\
ca-CondMat & 23,134 & 93,439 & 173,361 &        0.125122        &       7.47E+08        &       11.65   \\
ca-GrQc & 5,243 & 14,484 & 48,260 &     0.063396        &       2.28E+08        &       17.31   \\
ca-HepPh & 12,009 & 118,489 & 3,358,499  &      0.973678        &       1.22E+08        &       1.79    \\
ca-HepTh & 9,878 & 25,973 & 28,339 &    0.055146        &       4.71E+08        &       14.90   \\
cit-HepPh & 34,547 & 420,877 & 1,276,868 &      0.763726        &       5.51E+08        &       3.49    \\
cit-HepTh & 27,771 & 352,285 & 1,478,735 &      1.486301        &       2.37E+08        &       1.87    \\
cit-Patents & 3,774,769 & 16,518,947  &7,515,023 &      32.25143        &       5.12E+08        &       2.40    \\
email-Enron & 36,693 & 183,831 & 727,044  &     0.718355        &       2.56E+08        &       2.44    \\
email-EuAll & 265,215 & 364,481  & 267,313 &    1.946259        &       1.87E+08        &       1.84    \\
facebook\_combined & 4,040 & 88,234 & 1,612,010 &       1.030469        &       8.56E+07        &       1.39    \\
flickrEdges & 105,939 & 2,316,948 & 107,987,357 &       27.124476       &       8.54E+07        &       1.06    \\
graph500-scale18-ef16 & 174,148 & 3,800,348 & 82,287,285 &      1.0303421       &       1.69E+08        &       2.09    \\
graph500-scale19-ef16 & 335,319 & 7,729,675 & 186,288,972 &     3.23446107      &       1.04E+06        &   1.38    \\
graph500-scale20-ef16 & 645,821 & 15,680,861 & 419,349,784 &    9.46509552      &       6.82E+07        &       1.01    \\
graph500-scale21-ef16 & 1,243,073 & 31,731,650 & 935,100,883 &          29.26859307     &       4.25E+07        &       0.76    \\
loc-brightkite edges & 58,229 & 214,078 & 494,728 &     0.550747        &       3.89E+08        &       4.12    \\
loc-gowalla edges & 196,592 & 950,327 & 2,273,138 &     6.270409        &       1.52E+08        &       0.77    \\
oregon1\_010331 & 10,671 & 22,002 & 17,144 &    0.421596        &       5.22E+07        &       2.11    \\
oregon1\_010407 & 10,730 & 21,999 & 15,834 &    0.414824        &       5.30E+07        &       1.96    \\
oregon1\_010414 & 10,791 & 22,469 & 18,237  &   0.43478 &       5.17E+07        &       2.02    \\
oregon1\_010421 & 10,860 & 22,747 & 19,108 &    0.437284        &       5.20E+07        &       1.86    \\
oregon1\_010428 & 10,887 & 22,493 & 17,645 &    0.433493        &       5.19E+07        &       1.89    \\
oregon1\_010505 & 10,944 & 22,607 & 17,597 &    0.437427        &       5.17E+07        &       2.06    \\
oregon1\_010512 & 11,012 & 22,677 & 17,598 &    0.449109        &       5.05E+07        &       1.82    \\
oregon1\_010519 & 11,052 & 22,724 & 17,677 &    0.44961 &       5.05E+07        &       1.82    \\
oregon1\_010526 & 11,175 & 23,409 & 19,894 &    0.439644        &       5.32E+07        &       1.86    \\
oregon2\_010331 & 10,901 & 31,180 & 82,856  &   0.44663 &       6.98E+07        &       2.10    \\
oregon2\_010407 & 10,982 & 30,855 & 78,138  &   0.456381        &       6.76E+07        &       2.05    \\
oregon2\_010414 & 11,020 & 31,761 & 88,905  &   0.43776 &       7.26E+07        &       1.95    \\
oregon2\_010421 & 11,081 & 31,538 & 82,129 &    0.458956        &       6.87E+07        &       1.85    \\
oregon2\_010428 & 11,114 & 31,434 & 78,000 &    0.432467        &       7.27E+07        &       1.99    \\
oregon2\_010505 & 11,158  & 30,943 & 72,182 &   0.446486        &       6.93E+07        &       1.90    \\
oregon2\_010512 & 11,261 & 31,303 & 72,866 &    0.437808        &       7.15E+07        &       2.10    \\
oregon2\_010519 & 11,376 & 32,287 & 83,709 &    0.446916        &       7.22E+07        &       1.91    \\
oregon2\_010526 & 11,462 & 32,730 & 89,541 &    0.447869        &       7.31E+07        &       1.91    \\
p2p-Gnutella04  & 10,877 & 39,994 & 934 &       0.054383        &       7.35E+08        &       21.63   \\
p2p-Gnutella05 & 8,847 & 31,839 & 1,112 &       0.058627        &       5.43E+08        &       14.33   \\
p2p-Gnutella06 & 8,718 & 31,525 & 1,142 &       0.062394        &       5.05E+08        &       13.51   \\
p2p-Gnutella08 & 6,302 & 20,777 & 2,383 &       0.060773        &       3.42E+08        &       18.19   \\
p2p-Gnutella09 & 8,115 & 26,013 & 2,354 &       0.059509        &       4.37E+08        &       20.33   \\
p2p-Gnutella24 & 26,519 & 65,369 & 986 &        0.093102        &       7.02E+08        &       14.01   \\
p2p-Gnutella25 & 22,688 & 54,705 & 806 &        0.048971        &       1.12E+09        &       30.94   \\
p2p-Gnutella30 & 36,683 & 88,328 & 1,590 &      0.055981        &       1.58E+09        &       18.87   \\
p2p-Gnutella31 & 62,587 & 147,892 & 2,024 &     0.077367        &       1.91E+09        &       25.42   \\
roadNet-CA  & 1,965,207 & 2,766,607 & 120,676 &         0.282073        &       9.81E+09        &       49.29   \\
roadNet-PA & 1,088,093 & 1,541,898  & 67,150  &         0.180125        &       8.56E+09        &       38.56   \\
roadNet-TX & 1,379,918  & 1,921,660 & 82,869 &          0.20709 &       9.28E+09        &       41.61   \\
soc-Epinions1 & 75,880 & 405,740 & 1,624,481 &          2.308416        &       1.76E+08        &       1.35    \\
soc-Slashdot0811 & 77,361 & 469,180 & 551,724 &         2.095413        &       2.24E+08        &       2.25    \\
soc-Slashdot0902 & 82,169 & 504,230 & 602,592 &         2.20871 &       2.28E+08        &       1.51    \\
\bottomrule
\end{tabular}
\caption{Runtime (ms) and throughput (TEPS) for provided graphs. Speedup is over Hu et al.'s~\cite{Hu:2018:HTC} GraphChallenge submission last year (2018).
\label{tab:perf}
}
\end{table*}

\section{Conclusion}
\label{sec:conc}
This work is an update and deep dive on previous triangle counting work using subgraph matching approach~\cite{Wang:2016:ACS}. We improved both the filtering and joining phase by pruning more invalid nodes based on neighorhood encoding information, and using optimizations like $k$-step look-ahead to reduce unwanted intermediate results. We believe our optimizations are not limited to only triangle counting on the GPU\@. We expect the generality of our implementation allows others to extend this method to match more complicated subgraph patterns.

\section{Acknowledgement}
\label{sec:acks}
We appreciate the funding support from the Defense Advanced Research Projects Agency (Awards \# FA8650-18-2-7835 and HR0011-18-3-0007) and the National Science Foundation (Awards \# OAC-1740333 and CCF-1629657). This research was, in part, funded by the U.S. Government. The views and conclusions contained in this document are those of the authors and should not be interpreted as representing the official policies, either expressed or implied, of the U.S. Government. The first author, an NVIDIA Fellowship Finalist, thanks NVIDIA for research guidance. Also thanks to NVIDIA for equipment donations and server time.

\bibliographystyle{IEEEtran}
\bibliography{sm.bib}
%

\end{document}